\documentclass[prb,twocolumn,superscriptaddress,showpacs]{revtex4}

\usepackage{graphicx}
\usepackage{dcolumn}
\usepackage{amssymb}
\usepackage{bm}
\usepackage{epsfig}

\begin{document}
\bibliographystyle{apsrev}

\title{Multiple Andreev reflections in a quantum dot coupled to superconductors: \\Effects of  spin-orbit coupling}

\author{Fabrizio Dolcini}
\affiliation{Scuola Normale Superiore di Pisa and   NEST CNR-INFM, I-56126 Pisa, Italy}
\email{f.dolcini@sns.it}

\author{Luca Dell'Anna}
\affiliation{Institut f\"ur Theoretische Physik, Heinrich-Heine Universit\"at D\"usseldorf, D-40225 D\"usseldorf, Germany}
\affiliation{Scuola Internazionale Superiore  di Studi   Avanzati, via Beirut 2-4, I-34014, Trieste, Italy}
\email{dellanna@sissa.it} 
\date{\today}

\begin{abstract}
We study the out-of-equilibrium current through a multilevel quantum dot contacted to two superconducting leads and in the presence of Rashba and Dresselhaus spin-orbit couplings, in the regime of strong dot-lead coupling. The multiple Andreev reflection (MAR) subgap peaks in the current voltage characteristics are found to be modified (but not suppressed) by the spin-orbit interaction, in a way that strongly depends on the shape of the dot confining potential. In a perfectly isotropic dot the MAR peaks are enhanced when the strength $\alpha_R$ and $\alpha_D$ of Rashba and Dresselhaus terms are equal. When the  anisotropy of the dot confining potential increases the dependence of the subgap structure on the spin-orbit angle $\theta=\arctan(\alpha_D/\alpha_R)$ decreases. Furthermore, when an in-plane magnetic field is applied to a strongly anisotropic dot, the peaks of the non-linear conductance oscillate as a function of the magnetic field angle, and the location of the maxima and minima allows for a straightforward read-out of the spin-orbit angle $\theta$.
\end{abstract}

\pacs{73.23.-b, 74.50.+r, 74.45.+c,  71.70.Ej}

\maketitle
\section{Introduction}
Spin-orbit (SO) interaction is one of the most striking relativistic effects that can be observed in solid state systems. The seminal proposal of a spin-polarized field effect transistor by Datta and Das\cite{Datta-Das}, and the range of possible applications in spintronics\cite{spintronics} led to a remarkable progress in controlling\cite{Nitta-Taka} and measuring\cite{Luo,schaepers,russi} not only  the overall spin-orbit coupling constant\cite{Samuelson} but also the Dresselhaus and Rashba ratio\cite{Ganichev,Giglberger,Ensslin}. Recently an increasing interest has been devoted to studying the interplay between spin-orbit (SO) coupling and superconductivity. In particular many studies\cite{Bezuglyi, Krive, Nazarov, Dimitrova, Luca, Beri, Yang} have widely analyzed the influence of SO interaction on the equilibrium supercurrent (DC Josephson effect), pointing out that the multi-level nature of the dot is a crucial ingredient. Indeed SO coupling induces, in the presence of many levels, inter-band spin-flip processes which can affect the Josephson current in a non trivial way.

\begin{figure} 
\epsfig{file=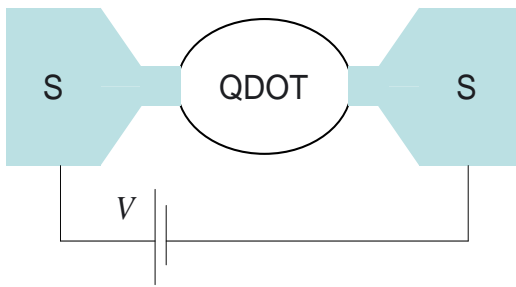,width=7cm,height=4cm,clip=}
\caption{\label{setup} (Color online) Scheme of the set-up under investigation: a quantum dot is coupled to two superconducting electrodes biased by a voltage $V$.}
\end{figure}

On the other hand, also {\it non-equilibrium} transport properties of nanodevices contacted to two superconducting electrodes have been a subject of intensive research since many years, from both theoretical \cite{Andreev, Artemenko, Blonder, Klapwijk, Arnold, Bratus, Averin, Alfredo} and experimental \cite{Ralph} sides. In particular, semiconductor-superconductor hybrid systems are currently under spotlight, because techniques have been developed to realize highly transparent semi/super interfaces\cite{GIA}, by suppressing Schottky barriers and enhancing the measured signal. In many cases such devices are based on In-As-related materials\cite{Nitta, van-Dam}, which are known to exhibit Rashba and Dresselhaus SO interaction. Nevertheless, a systematic analysis of the interplay between finite bias superconducting transport and spin precession  induced by SO is lacking. The aim of the present work is to bridge this gap, investigating the effects of SO on electron transport through a quantum dot (QD) connected to two superconducting leads, biased by a voltage $V$, as sketched in Fig.\ref{setup}.\\

It is  well known  that   the subgap I-V curve of transport through a quantum dot is characterized by Andreev peaks, originating from the resonance of the dot level with a  multiple Andreev reflection (MAR)  trajectory, i.e. a sequence of odd Andreev reflections occurring at the interfaces with the leads\cite{Alfredo}. In each Andreev reflection  the incoming and outgoing electron-hole pairs have opposite spins\cite{Blonder,Klapwijk}. In the presence of spin-orbit, however, the dot levels lack of a definite spin orientation. The question thus arises whether the resonances with dot levels and Andreev processes persist, or whether spin-orbit coupling suppresses the MAR peaks. Another interesting issue is the role of the shape of the dot confining potential; its anisotropies indeed affect not only the number of effective levels involved in transport, but also the relative weight of Rashba and Dresselhaus components. 
The scenario is expected to be even richer in the presence of an in-plane magnetic field, since the latter introduces a preferred spin direction competing with the spin-orbit precession. \\

Here we analyze how the subgap MAR pattern is affected by these phenomena. Our approach is based on the Keldysh technique, which allows us to account for non-equilibrium regime induced by the finite voltage. In order to single out the effects of spin-orbit interaction, we neglect  Coulomb interaction in the dot, so that tunneling between the leads and the dot can be treated exactly. Our results thus apply to the regime of relatively high dot-lead transparency (the linewidths are assumed to be some fraction of the superconducting gap), so that  spurious effects due to charging are avoided. We also consider different shapes for the dot confinement potential, and analyze the role of anisotropy.\\
The structure of the paper is the following. In Sec.\ref{SecII} we describe the model. The expression for the current is presented and briefly outlined in Sec.\ref{SecIII}. Results are then presented in Sec.\ref{SecIV}, which is divided into three subsections, addressing the effects of spin-orbit on the subgap structure for an isotropic dot, the role of anisotropy and of an in-plane magnetic field respectively. Conclusions can be found in the last section V, and details of the calculation in the Appendices. 
\section{The model}
\label{SecII}
We start by describing the  isolated dot. The Hamiltonian reads\cite{Luca}
\begin{eqnarray}
\lefteqn{\mathcal{H}_{D} =} & & \label{Ham-D} \\
& =& \! \int \! \! d\mathbf{r} \,  
d^\dagger(\mathbf{r})
\left[ 
\frac{\mathbf{p}^2}{2m^*}  +V_0(\mathbf{r})+ H_{magn} + V_{\rm SO}  \right] \! d^{}(\mathbf{r}) \nonumber
\end{eqnarray}
where $d(\mathbf{r})=(d_\uparrow(\mathbf{r}),d_\downarrow(\mathbf{r}))^T$ is the two spin component electron field operator in the dot. The first term in the brackets of Eq.~(\ref{Ham-D}) describes the kinetic energy, $\mathbf{p}=-i \hbar \nabla$ being the momentum operator and $m^*$ the electron effective mass. The second term 
\begin{equation}
V_0(\mathbf{r})=V_\parallel(x,y)
+V_\perp(z) \label{V0}
\end{equation}
models the dot confining potential, for which we have assumed a harmonic confinement 
\begin{equation}
V_\parallel(x,y)=
\frac{m^*}{2} (\omega_x x^2 + \omega_y y^2)   \label{Vpar}
\end{equation}
with frequencies $\omega_{x,y}$  in the 2DEG $x-y$-plane, and a hard wall potential $V_\perp$ along the growth direction $z$.  The third term
\begin{eqnarray}
H_{magn} &=& g  \mu_B \vec{H} \cdot  \frac{\vec{\sigma}}{2} = \label{Hmagn} \\
&=& \frac{g \mu_B H}{2} (\sigma_x \cos \phi_H + \sigma_y \sin \phi_H ) \nonumber
\end{eqnarray}
describes the coupling with an in-plane magnetic field $\vec{H}$ with intensity $H$ and angle $\phi_H$ with the $x$ axis. In Eq.~(\ref{Hmagn})  $g$ is the gyromagnetic factor, $\mu_B$ the Bohr magneton, and $\sigma_x,\sigma_y$ the Pauli matrices. Finally 
\begin{eqnarray}
V_{\rm SO} = \frac{\alpha_R}{\hbar}  \left( \sigma_x p_y -\sigma_y p_x \right) +\frac{\alpha_D}{\hbar}  \left( \sigma_x p_x -\sigma_y p_y \right)  \label{VSO}
\end{eqnarray}
is the spin-orbit coupling  accounting both for Rashba and Dresselhaus components with coupling constants $\alpha_R$ and $\alpha_D$ respectively. \\ 

We adopt as a basis the orbital eigenstates $\chi_n(\mathbf{r})$ of the dot in the absence of spin-orbit coupling and magnetic field
\begin{equation}
\left( -\frac{ \hbar^2 \nabla^2}{2m^*}   +V_0(\mathbf{r}) \right) \chi_n(\mathbf{r})= \varepsilon^0_n \, \chi_n(\mathbf{r})
\end{equation}
where $\varepsilon^0_n$ are the related eigenvalues, measured with respect to the equilibrium Fermi energy $E_F$.  Typically transport properties within a voltage range comparable to $|\Delta|$ only involve a few dot levels. We assume that the width in the growth direction $z$ is small, i.e. that the level spacing due to $z$ confinement is large compared to $|\Delta|$, so that only the lowest state along $z$ is involved and that the electron dynamics is actually two-dimensional. Thus one has
\begin{equation}
\varepsilon^0_{n_x,n_y}=E_0+ \hbar (\omega_x n_x+\omega_y n_y) \hspace{1cm}n_x,n_y=0,1 \ldots \label{spectrum0}
\end{equation}
with $E_0$ denoting the ground state level. The levels contributing to transport thus depend on the ratio of the confining energies $\hbar \omega_{x,y}$ with respect to $|\Delta|$. Denoting by $N$ the number of these effective orbital levels, only $2N$ states determine the current-voltage characteristics (the factor 2 arising from spin), and the electron field can fairly be  approximated as
\begin{equation}
d_\sigma(\mathbf{r}) \simeq \sum_{n=1}^N \chi_n(\mathbf{r}) \, d_{n  \sigma} \label{field-approx}
\end{equation}
where   $d^\dagger_{n,\sigma}$ and $d_{n \sigma}$ ($n=1,\ldots N$) denote the related creation and annihilation operators, respectively. Inserting Eq.~(\ref{field-approx}) into Eq.~(\ref{Ham-D}), one obtains a $2N \times 2N$ matrix representation $H_D$ for the dot Hamiltonian  
\begin{equation}
\mathcal{H}_D= \left( 
d^\dagger_{1 \uparrow}, 
\ldots ,
d^\dagger_{N \uparrow},
d^\dagger_{1 \downarrow} ,
\ldots ,
d^\dagger_{N,\downarrow} 
\right) \! \cdot  
 H_D  \cdot \! \left( 
\begin{array}{c} 
d_{1 \uparrow} \\ 
\vdots \\
d_{N \uparrow}  \\
d_{1 \downarrow} \\ 
\vdots \\
d_{N \downarrow} 
\end{array}
\right)  \label{HD-mat}
\end{equation}
which can be diagonalized.
\\For the leads we adopt the customary 3D s-wave BCS model. Labeling by $p=+$ ($p=-$) the  left (right) lead, one has
\begin{eqnarray}
\lefteqn{ \mathcal{H}_{lead,p} = \! \!  \sum_{\mathbf{k},\sigma=\uparrow,\downarrow} \xi^{}_{\mathbf{k}} \, c^\dagger_{\mathbf{k}  p  \sigma} c^{}_{\mathbf{k}  p  \sigma} +} & & \label{HBCS} \\
& & \hspace{1cm} + \sum_{\mathbf{k}} (|\Delta| e^{i p\phi/2}  c^\dagger_{\mathbf{k}  p  \uparrow} c^\dagger_{\mathbf{k}  p  \downarrow} + \mbox{h.c.} ) \nonumber 
\end{eqnarray}
where $c^\dagger_{\mathbf{k}  p  \sigma}$ denotes the electron creation operators with momentum $\mathbf{k}$ in the $p$-th lead and with spin projection $\sigma=\uparrow,\downarrow$, $|\Delta|$  the half-gap (supposed the same in both electrodes), $\phi$ the equilibrium superconducting phase difference between the two leads, and $\xi_{\mathbf{k}}=\mathbf{k}^2/2m-E_F$ the excitation energy with respect to the Fermi level $E_F$. Here $E_F$ is the Fermi level of the two leads at equilibrium; the applied bias $V$ is included as a time-dependent tunneling amplitude (see below).\\

Finally, a term 
\begin{eqnarray}
\mathcal{H}_{tun,p} &=&   \sum_{\sigma=\uparrow,\downarrow} \sum_{\mathbf{k}} \sum_{n=1}^N 
\left( 
t_{p,n}(t)   c^\dagger_{\mathbf{k}, p, \sigma} d^{}_{n \sigma}  +  \mbox{h.c.}
\right)  \label{Htun}
\end{eqnarray}
describes the tunneling between dot and leads. We have assumed for simplicity that the tunneling amplitudes $t_{p,n}$ between the $p$-th lead and the level $n$ of the dot are spin and $\mathbf{k}$-independent. A generalization is straightforward. The explicit time-dependence $t_{p,n}(t)=t_{p,n}(0) \exp (i p \omega_V t/2)$ accounts for the bias $V$ applied between the superconductors\cite{Alfredo}, where 
\begin{equation}
\omega_V=\frac{e V}{\hbar} 
\end{equation}
is the frequency related to the bias. The   energies $\Gamma_{p,n}=\pi \rho(\varepsilon_F) |t_{p,n}(0)|^2$ are the  tunneling linewidths associated with tunneling from the $n$-th dot level into the $p$-the lead, with $\rho(\varepsilon_F)$ being the DOS in the normal state.\\

In conclusion the total Hamiltonian of the system reads
\begin{equation}
\mathcal{H}= \mathcal{H}_{D}+ \mathcal{H}_{lead,+} + \mathcal{H}_{lead,-} +  \mathcal{H}_{tun,+} + \mathcal{H}_{tun,-} \label{Ham} 
\end{equation}
where $\mathcal{H}_{D}$, $\mathcal{H}_{lead,\pm}$, and $\mathcal{H}_{tun,p}$ are given by Eqs.(\ref{Ham-D}), (\ref{HBCS}) and (\ref{Htun}) respectively. 
\section{Current}
\label{SecIII}
The current flowing from $p$-th lead into the dot can be written as
\begin{eqnarray}
I_p(t) = \frac{ie}{\hbar} \sum_{\sigma=\uparrow,\downarrow} \sum_{\mathbf{k}} \sum_{n=1}^N 
\left( 
t_{p,n}(t) \langle c^\dagger_{p, \sigma} d^{}_{n \sigma} \rangle- {\rm h.c.}  
\right) \label{Ip-prel}
\end{eqnarray}
Combining the non-equilibrium Green function technique with the Dyson equation, one can rewrite Eq.~(\ref{Ip-prel}) as
\begin{eqnarray}
I_p(t)= - 2e   \Re \! \int_{-\infty}^{\infty} \! \! \! \! \! \! dt^\prime {\rm Tr}_{4N} \left\{   (\sigma_z)_{4N} \left[ \mathsf{\Sigma}_p(t,t^\prime)   \mathsf{G}(t^\prime,t)  \right]^{+-} \! \right\}    \label{Ip-fin-1} 
\end{eqnarray}
Here bold notations denote matrices in the Keldysh   space of the quantum dot, with the superscript $+$($-$) labeling the upper (lower) Keldysh time contour branch. In particular  $\mathsf{\Sigma}_p$ and $\mathsf{G}$ respectively describe the self-energy due to $p$-th lead and the dot Green function, evaluated in the presence of the leads
\begin{equation}
\mathsf{G}=(\mathsf{G}_0^{-1}-\mathsf{\Sigma}_{+}-\mathsf{\Sigma}_{-})^{-1} \label{Dyson-2}
\end{equation}
where $\mathsf{G}_0$ describes the isolated dot.     Details of the derivation  of Eq.~(\ref{Ip-fin-1})  can be found in App.\ref{AppA}. Here we  emphasize that, differently from the customary treatment of superconducting transport through a quantum dot, in our case one has to adopt a Nambu space for the dot with dimension $4N$, instead of $2N$, due to the spin-orbit term which effectively induces spin-flip tunneling processes. For these reasons in Eq.~(\ref{Ip-fin-1}) the symbol ${\rm Tr}_{4N}$ denotes the trace on the $4N$-dimensional (extended) Nambu space of the dot, and $(\sigma_z)_{4N}= \sigma_z \otimes {\mathbb{I}}_{2N}$, where $\sigma_z$ is the usual Pauli matrix and  ${\mathbb{I}}_{2N}$ the $2N \times 2N$ identity matrix.\\

The presence of a finite bias yields the current to be time-dependent (AC Josephson effect); it is thus suitable to derive a harmonic time series representation of it. By applying a discrete Fourier Transform (see App.\ref{AppB}) to $\mathsf{\Sigma}_p(t,t^\prime)$ and $\mathsf{G}(t^\prime,t)$, one obtains
\begin{eqnarray}
\lefteqn{I_p(t) = 
- \frac{e}{\pi}  \,  \Re \sum_{m=-\infty}^\infty  \, e^{+i m \omega_V t}  \, \int_F  d\omega  \, {\rm Tr}_{4N}   (\sigma_z)_{4N}   } & &   \nonumber   \\
& &  \times \left[  \sum_{n_1,n_2=-\infty}^\infty \mathsf{\Sigma}_{p}(n_1,n_2;\omega) \, \mathsf{G}(n_2,n_1+m;\omega)    \,    \right]^{+-}     \label{Ip-fin}
\end{eqnarray}
where 
\begin{equation}
F=[-\frac{\omega_V}{2} \, , \, \frac{\omega_V}{2}] \label{FUND-DOM}
\end{equation}
is the voltage-dependent fundamental domain, and $m$ labels the harmonic. 
The Fourier coefficients $\mathsf{\Sigma}_{p}(n_1,n_2;\omega)$ and $\mathsf{G}(n_1,n_2;\omega)$ of the self energy  and the dot Green function 
   are computed in App.\ref{AppB}. We emphasize that, since Eq.~(\ref{Ip-fin}) is formally exact, charge conservation implies that
\begin{equation}
I_{+}(t)=-I_{-}(t) \doteq I(t) \label{conserv}
\end{equation}
In the explicit calculation the series in $n_1$ and $n_2$ must be  truncated to a cut-off, which is chosen in order to ensure convergence of $I_{\pm}$ and the fulfillment of Eq.~(\ref{conserv}).\\

The DC component of the current (\ref{conserv}) corresponds to the $m=0$ harmonic and thus reads
\begin{eqnarray}
\lefteqn{I_0 = 
-p \frac{e}{\pi}  \,  \Re   \, \int_F  d\omega  \, {\rm Tr}_{4N}   (\sigma_z)_{4N}   } & &   \nonumber   \\
& &  \times \left[  \sum_{n_1,n_2=-\infty}^\infty \mathsf{\Sigma}_{p}(n_1,n_2;\omega) \, \mathsf{G}(n_2,n_1;\omega)    \,    \right]^{+-}     \label{Ip-fin-DC}
\end{eqnarray}
where $p=\pm 1$. While the case of equilibrium Josephson current ($V=0$ and $\phi \neq 0$) was thoroughly analyzed in Ref.\onlinecite{Luca}, here we focus on the out of equilibrium case ($V \neq 0$), and henceforth set $\phi=0$. The current (\ref{Ip-fin-DC}) is then evaluated with numerical integration in the frequency. 
\section{Spin-orbit and Multiple Andreev reflections}
\label{SecIV}
\begin{figure} 
\epsfig{file=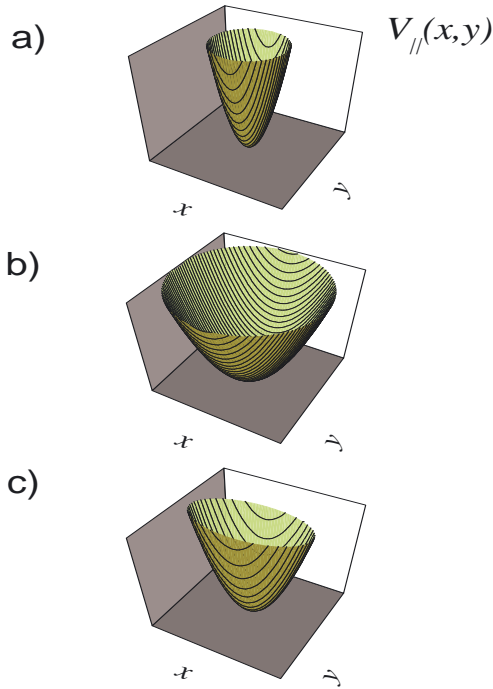,width=7cm,height=10cm,clip=}
\caption{\label{parabolas} (Color online) Sketch of the dot confining potential (\ref{Vpar}) for different values of the frequencies in the $x$ and $y$ directions.  a)  $\hbar \omega_{x,y} \gg |\Delta|$ : only one level takes part to transport; b)  $\hbar \omega_{x} = \hbar \omega_{y} \simeq |\Delta|$: multilevel isotropic dot; c) $\hbar \omega_{x} \simeq |\Delta|$ and $\hbar \omega_{y} \gg |\Delta|$: multilevel strongly anisotropic dot.}
\end{figure}
Here we present the results concerning the DC component (\ref{Ip-fin-DC}) of the current, analyzing the effects of spin-orbit interaction on the multiple Andreev reflections pattern and subgap structure. We consider  different types of shapes for the  confining potential $V_\parallel$, from isotropic ($\omega_{x}=\omega_y$) to strongly anisotropic ($\omega_{y} \gg \omega_x$), as shown in Fig.\ref{parabolas}. Experimentally, a tuning of the frequencies $\omega_{x,y}$ (see Eq.~(\ref{Vpar})) can be achieved  through side gates, which do not substantially alter Rashba and Dresselhaus spin-orbit coupling. For the sake of clarity, it is worth pointing out that throughout the paper we use the term isotropy as referring to the confining potential $V_\parallel$ only, and not to the full dot Hamiltonian. It is indeed easily checked that, even for a isotropic potential $V_\parallel$ and in   absence of in-plane magnetic field, the Hamiltonian of the dot is {\it not} invariant under rotation around $z$-axis, because the Dresselhaus term in Eq.~(\ref{VSO}) does not commute with the total angular momentum $J_z=L_z+S_z$. \\ 

We start by some general remarks.
In the first instance, simple perturbation theory arguments lead to conclude that spin-orbit effects can only be observed in a multilevel quantum dot:  for one single level quantum dot, spin-orbit interaction does not play any role, because the momentum operator appearing in Eq.~(\ref{VSO}) can only couple dot levels with different  quantum numbers $n_x$ and $n_y$ (see Eq.~(\ref{spectrum0}))\cite{NOTA}. For these reasons, in the following we shall consider only multilevel quantum dots. The number $2N$ of states contributing to transport depends on the parameters of $V_{\parallel}(x,y)$, and   is chosen in such a way that the subgap I-V curve remains unaffected by inclusion of higher energy levels. Notice that for intermediate level spacing (i.e. for $\hbar \omega_{x,y} > |\Delta|$) the dot excited levels are energetically too high to contribute to transport directly, but close enough to the unperturbed ground state to modify its resonance conditions via spin-orbit interaction, hence affecting transport indirectly. In this sense the effect of spin orbit can be considered as due to a single level.\\

Secondly,   the shape of the confining potential has a crucial role in determining spin-orbit effects: it is easy to verify that  entries  of the dot Hamiltonian (\ref{Ham-D})   originating from spin-orbit couplings (\ref{VSO}) scale as $\alpha_{R,D}/\lambda_{x,y}$, where 
\begin{equation}
\lambda_{x,y} = \sqrt{\frac{\hbar}{m^* \omega_{x,y}}}
\end{equation}
are the lengths associated with the frequencies of the confining potential $V_\parallel$. Indeed the value of $\omega_x/\omega_y$ determines the number of states in the $x$ and $y$ direction that contribute to transport, and also changes the relative weight of spin-orbit terms proportional to $p_x$ with respect to those involving $p_y$.\\

Finally, since we account for both Rashba and Dresselhaus terms, it is worth introducing a total Dresselhaus/Rashba coupling constant\cite{Luca}
\begin{equation}
\alpha=\sqrt{\alpha_D^2+\alpha_R^2} \label{alpha}
\end{equation}
and a relative Dresselhaus/Rashba spin-orbit angle $\theta$, defined through the relations
\begin{equation}
\cos\theta=\frac{\alpha_R}{\alpha} \hspace{2cm} \sin\theta =\frac{\alpha_D}{\alpha} . \label{thetaSO}
\end{equation} 
With the help of Eqs.(\ref{alpha}) and (\ref{thetaSO}), the spin-orbit coupling (\ref{VSO}) can be rewritten as
\begin{eqnarray}
V_{\rm SO} = \frac{\alpha}{\hbar} \, \vec{p}_\parallel \cdot \vec{a}  \label{VSO-2}
\end{eqnarray}
where $\vec{p}_\parallel=(p_x,p_y)$ is the $x-y$ plane momentum and $\vec{a}=(a_x,a_y)$ is a vector with components
\begin{eqnarray} 
a_x  &=& \sigma_x \sin \theta   -\sigma_y \cos\theta  \\
a_y  &=& \sigma_x \cos \theta   -\sigma_y \sin\theta  \quad.
\end{eqnarray}
Eq.~(\ref{VSO-2}) describes the well known effect that, in the presence of spin-orbit coupling,  the spin orientation depends on the momentum $\vec{p}_\parallel$ direction, in a manner which is also related Rashba/Dresselhaus angle $\theta$. For the case of a two-dimensional electron gas, this aspect was recently emphasized e.g. in Ref.\onlinecite{Giglberger}. In a quantum dot, however, states with well defined momentum are not eigenstates, due to the presence of the confining potential~$V_\parallel$. Hence, in the presence of spin-orbit coupling the dot levels do not exhibit a definite spin orientation in general, and the question arises whether MAR resonances are suppressed by spin-orbit.  \\

\begin{figure} 
\epsfig{file=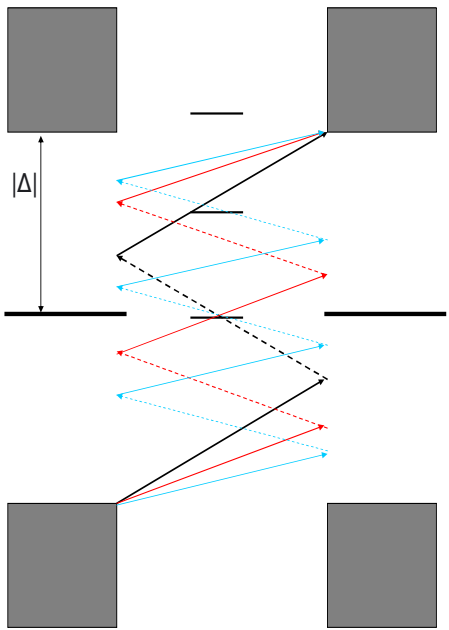,width=6cm,height=8.5cm,clip=}
\caption{\label{Fig-MAR} (Color online) Pictorial scheme of multiple Andreev reflections in a multilevel quantum dot. The left and right parts represent the energy spectrum of the superconducting leads, characterized by a gap $|\Delta|$, whereas the horizontal lines in the central part of the figure describe the dot levels. The subgap MAR processes related to $eV/2 |\Delta|=1/n$ with $n=3,5,7$ are depicted.}
\end{figure}

In the rest of the paper we present our results addressing  the following questions: i) does spin-orbit interaction in a quantum dot suppress MAR peaks? (see Fig.\ref{Fig-MAR}) ii) does the geometry of the dot matter in observing spin-orbit effect on electron transport; iii) what is the effect of the interplay between the spin-orbit angle $\theta$ and the magnetic field on the I-V curves? \\
\subsection{Spin-orbit effects for an isotropic dot}
\begin{figure} 
\epsfig{file=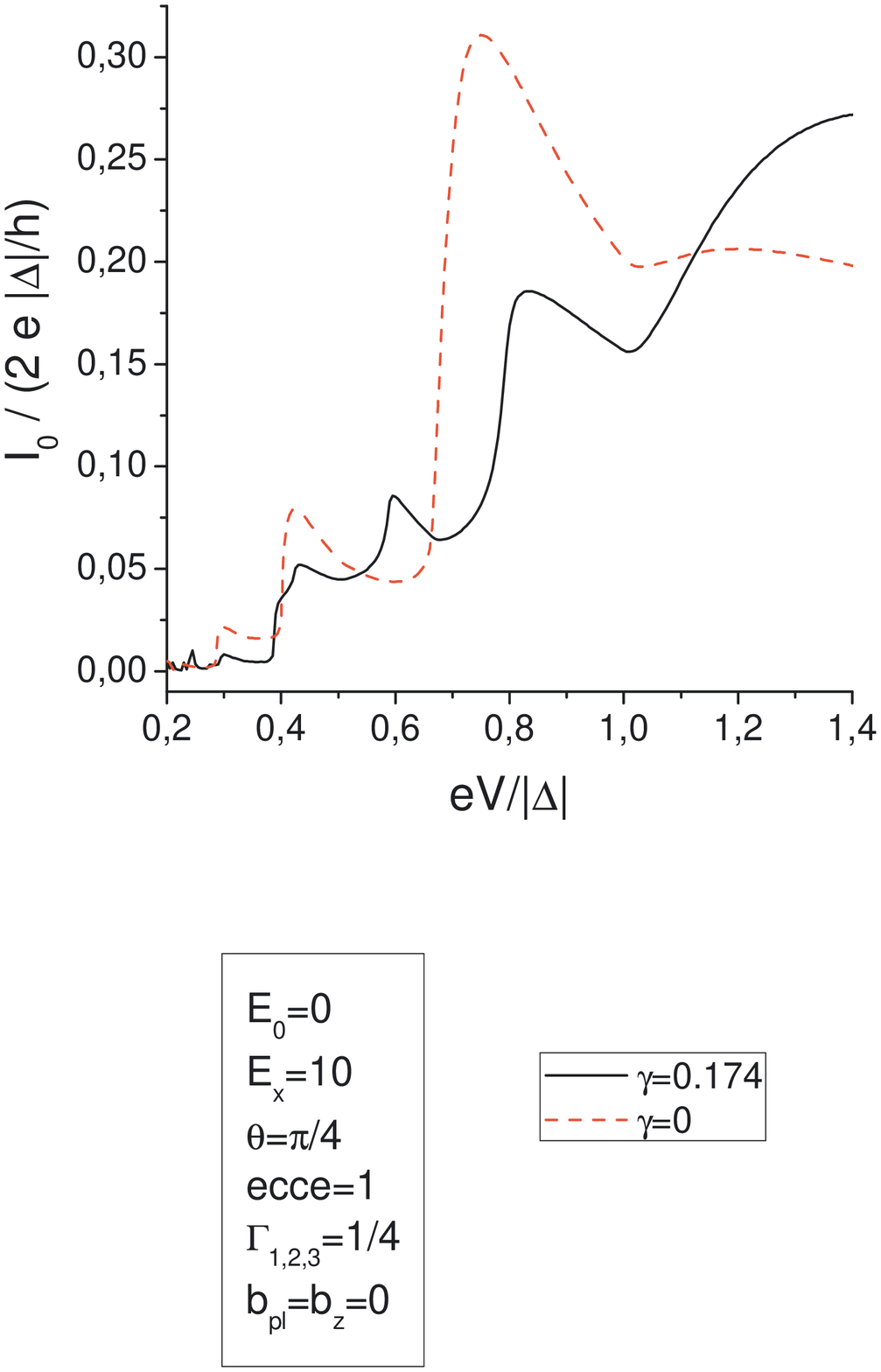,width=7cm,height=7cm,clip=}
\caption{\label{Figura-SYM-1} (Color online)
 Current (units of $2e|\Delta|/h$) as a function of the source-drain bias for a three-level quantum dot with isotropic confining potential ($\hbar \omega_x=\hbar \omega_y=20 |\Delta|$) and linewidth $\Gamma_{\pm,i}=|\Delta|/4$ ($i=1,2,3$). The solid curve refers to the case of spin-orbit coupling $\alpha = 10^{-11} eV m$ ($\gamma=0.174$) and spin-orbit angle $\theta=\pi/4$. The dashed curve describes the case of vanishing spin-orbit coupling $\alpha=0$. The spin-orbit coupling does not suppress  the MAR subgap pattern; it modifies the location and the number of the peaks.}
\end{figure}

We start our analysis from the case of an isotropic confining potential ($\omega_x=\omega_y$), without magnetic field. 
The solid curve of Fig.\ref{Figura-SYM-1} shows   the I-V characteristics for a dot with $E_0=0$ and confining potential frequencies $\hbar \omega_x=\hbar \omega_y= 20|\Delta|$, so that only the ground state lies in the gap energy window. The solid curve is obtained in the presence of spin-orbit coupling $\alpha=10^{-11} {\rm eVm}$ and Rashba/Dresselhaus angle $\theta=\pi/4$. For the sake of comparison we have also plotted the case of vanishing spin-orbit (dashed curve). Our result indicates that spin-orbit coupling modifies, rather than suppress, the MAR pattern. This is due to the fact that, although spin-orbit coupling does not allow electrons in the dot to have a definite spin orientation, it does not break time-reversal symmetry (TRS).
 As a consequence of TRS, the dot levels are always doubly degenerate and the related pair of eigenstates mutually connected by the time-reversal transformation
\begin{equation}
\mathsf{T}= i \sigma_y \mathsf{K} 
\end{equation}
where $\mathsf{K}$ is complex conjugation. 
An electron emerging from an Andreev reflection at the interface can thus tunnel into the doubly degenerate dot level, independent of its original spin-direction, even in the presence of a finite spin orbit coupling $\alpha$. With respect to the case $\alpha=0$, the   electron simply redistributes differently between the two TRS dot states. Although spin-orbit effect does not destroy the MAR subgap pattern, it does modify it, because the location of MAR peaks and their sharpness depend on the dot resonant levels and the dot-lead linewidths, which are both affected by spin-orbit coupling. 
The simple case of an isotropic quantum dot with 3 levels (one $s$ and two orbitally degenerate $p$ levels) can elucidate this effect, since the spectrum can be evaluated analytically, obtaining
\begin{eqnarray}
\varepsilon_1 &=& E_0+\hbar \omega_x \frac{1-\sqrt{1+2 \gamma^2}}{2}   \\
\varepsilon_2 &=& E_0+\hbar \omega_x   \\
\varepsilon_3 & =& E_0+ \hbar \omega_x \frac{1+\sqrt{1+2 \gamma^2}}{2}   
\end{eqnarray}
where
\begin{equation}
\gamma=\frac{\alpha}{\hbar} \sqrt{\frac{2m^*}{\hbar \omega_x}} \label{gamma}
\end{equation}
is the dimensionless spin-orbit coupling.\\

In conclusion, the  MAR pattern is   quantitatively modified by  spin orbit coupling $\alpha$, although it qualitatively resembles the one of a  dot with appropriately renormalized levels and linewidths.  \\

\begin{figure}
\epsfig{file=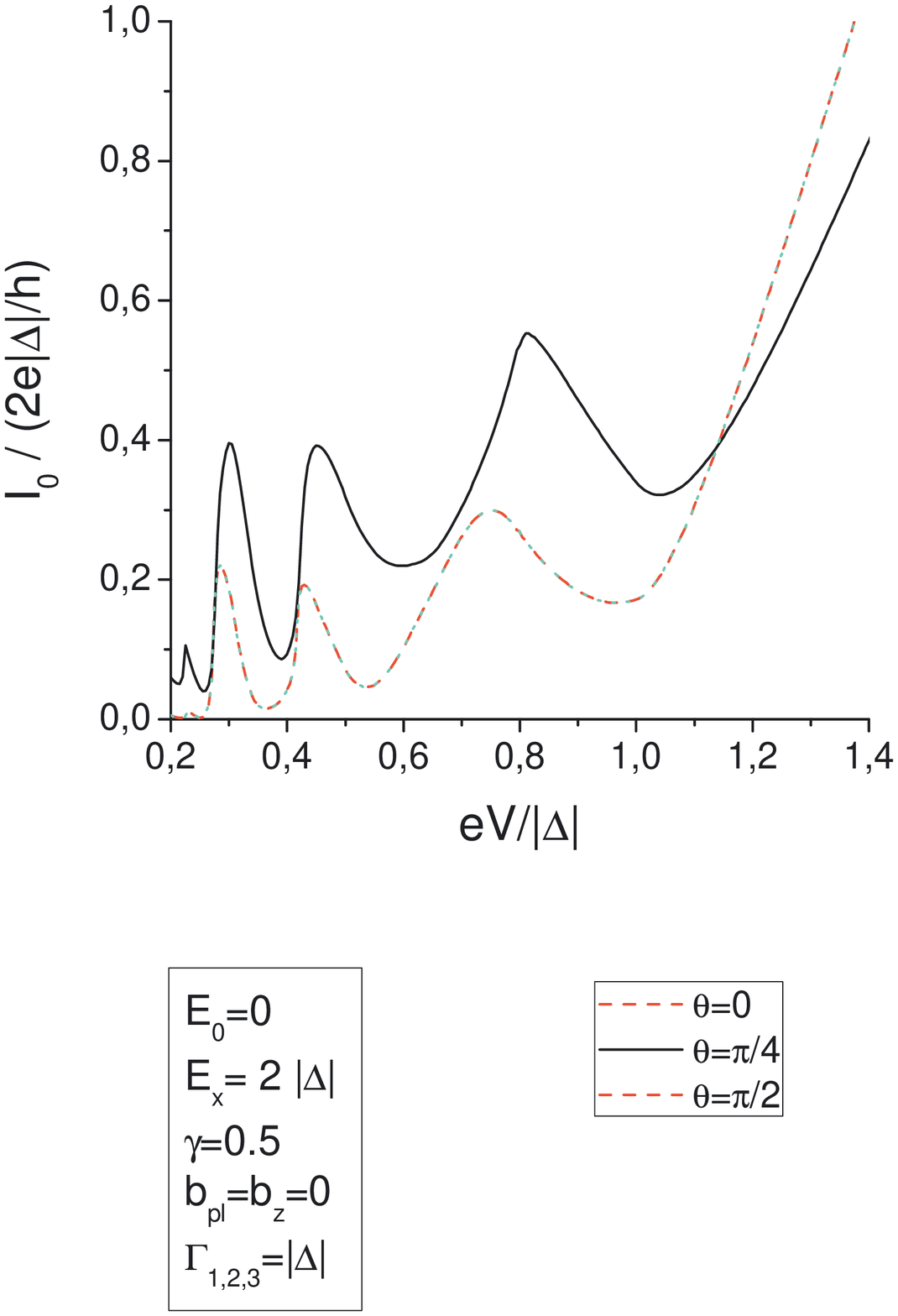,width=7cm,height=7.5cm,clip=}
\caption{\label{Figura-SYM-2} 
(Color online) Current-voltage characteristics   for a 3-level quantum dot with isotropic confining potential ($\hbar \omega_x=\hbar \omega_y=4 |\Delta|$) and linewidths $\Gamma_{\pm,i}=|\Delta|$ ($i=1,2,3$), in the presence of a spin-orbit coupling constant $\gamma=0.5$ (see Eq.~(\ref{gamma})) and for different values of the Rashba/Dresselhaus angle $\theta$. The solid curve refers to $\theta=\pi/4$, i.e. equal weight of Rashba and Dresselhaus terms, whereas the dashed (dotted) curve refers to $\theta=0$ ($\theta=\pi/2$), corresponding to purely Rashba (purely Dresselhaus) coupling. The mixed coupling enhances the MAR peaks, whereas the two other cases turn out to coincide, due to symmetry reasons (see text).}
\end{figure}

Let us now discuss the effects of the spin-orbit angle~$\theta$. In Fig.\ref{Figura-SYM-2} the current-voltage characteristics is shown for different values of $\theta$. One can notice that for an equal weight of Rashba and Dresselhaus terms ($\theta=\pi/4$) the MAR  peaks are enhanced with respect to the two cases of purely Rashba ($\theta=0$) and purely Dresselhaus ($\theta=\pi/2$) interactions. Indeed for the particular value $\theta=\pi/4$ electrons in the dot have a well defined spin orientation, since the Hamiltonian (\ref{Ham-D}) commutes with the spin operator $(\sigma_{x}-\sigma_y)/\sqrt{2}$. In contrast, for arbitrary values of $\theta$ spin is not a good quantum number. The curves at $\theta=0$ and $\theta=\pi/2$ turn out to coincide, since the Hamiltonian corresponding to these two cases can be  mapped into  each other by the unitary transformation on the spin variables
\begin{eqnarray}
\mathsf{V} : \left\{ \begin{array}{lcl} \sigma_x & \leftrightarrow & -\sigma_y \\
\sigma_z & \rightarrow & -\sigma_z   \end{array}\right. \quad.
\end{eqnarray}
\subsection{The effect of the anisotropy in the confining potential in the absence of magnetic field}
\begin{figure} 
\epsfig{file=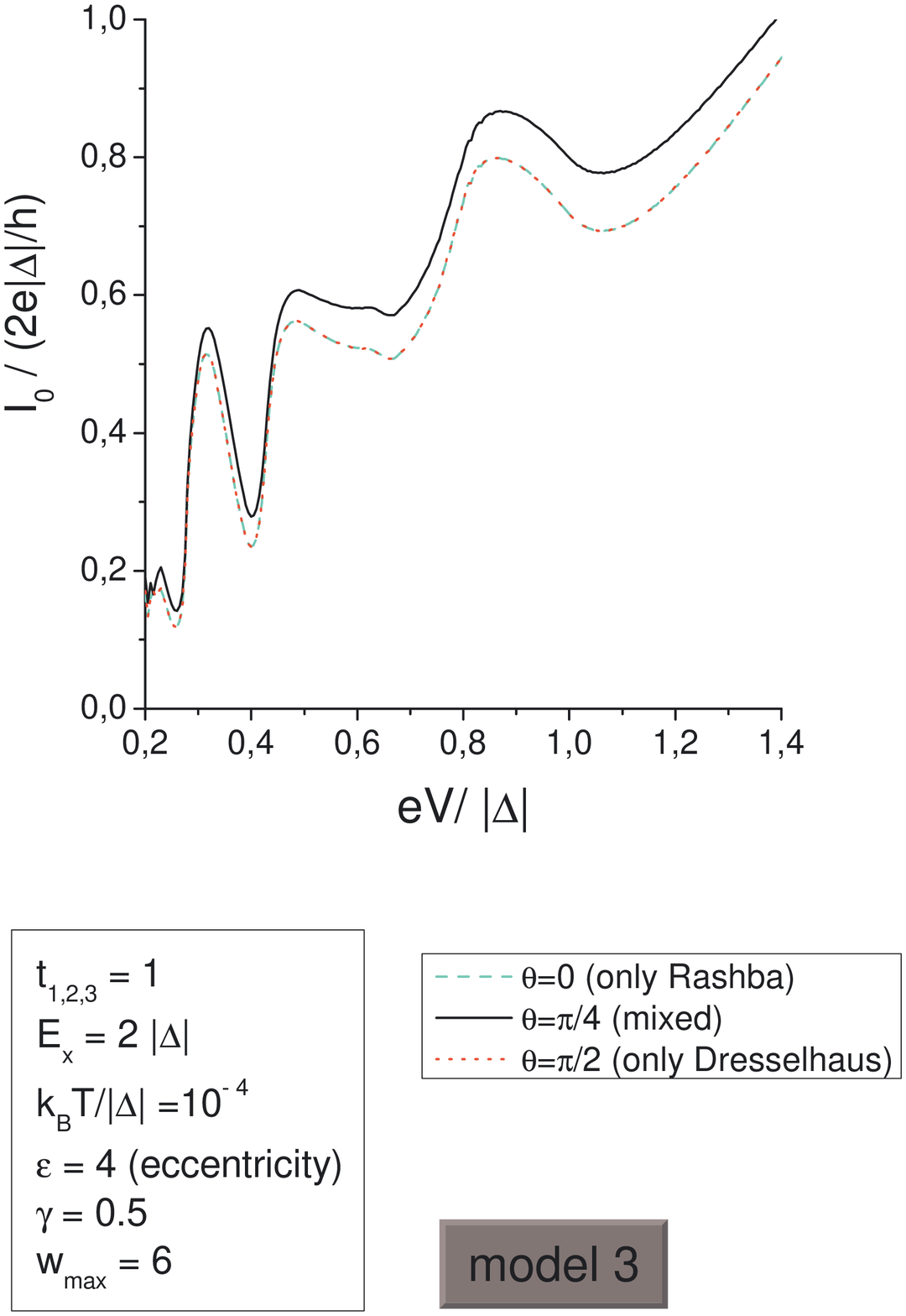,width=7cm,height=7.5cm,clip=}
\caption{\label{Figura-ASYM-1} (Color online) Current (units of $2e|\Delta|/h$) as a function of the source-drain bias, for a 3-level dot in an anisotropic confining potential ($\omega_y=16 \omega_x$), with dot-lead linewidths $\Gamma_{\pm,i}=|\Delta|$ ($i=1,2,3$) and spin-orbit coupling constant $\gamma=0.5$ (see Eq.~(\ref{gamma})), for different values of the Rashba/Dresselhaus angle $\theta$. The solid curve refers to $\theta=\pi/4$, i.e. equal weight of Rashba and Dresselhaus terms, whereas the dashed (dotted) curve refers to $\theta=0$ ($\theta=\pi/2$), corresponding to purely Rashba (purely Dresselhaus) coupling. With respect to the isotropic case, the dependence on $\theta$ is much weaker in the presence of an anisotropic confining potential.}
\end{figure}

The equivalence discussed above between transport properties of a dot with purely Rashba and purely Dresselhaus terms persists also in the presence of anisotropy in the confining potential $V_\parallel$ of the dot. However, when the anisotropy of the dot increases, the dependence of the I-V curves on the angle $\theta$ becomes weaker (see Fig.\ref{Figura-ASYM-1}). In particular, for a dot with a strongly anisotropic confining potential (e.g. for $\hbar \omega_y \gg \hbar \omega_x, |\Delta|$), and in absence of magnetic field, electron transport becomes insensitive to the spin-orbit angle $\theta$, and only the total intensity $\alpha$ of spin-orbit coupling matters. 
In order to understand this effect one can observe that, due to the anisotropy of the confining potential, only the lowest quantum number related to the $y$ effectively matters. Thus the operator~$p_y$ appearing in the spin-orbit interaction in Eq.~(\ref{VSO}) does not alter the dot levels involved in transport, and can effectively be dropped from the Hamiltonian. Importantly, this implies that for a strongly anisotropic quantum dot the spin-orbit interaction does not induce any spin-precession, but rather determines one preferred orientation of the electron spin in the $x-y$-plane. Although such spin direction formally depends on the spin-orbit angle $\theta$, the latter can indeed be gauged away completely from the Hamiltonian by performing a unitary transformation only on the spin degrees of freedom
\begin{equation}
\mathsf{V}_\theta= \exp\left[ i (\frac{\pi}{2}-\theta) \sigma_z/2 \right] \label{V-theta}
\end{equation}
yielding a spin-orbit coupling 
\begin{equation}
V_{SO} \rightarrow \frac{\alpha}{\hbar}   p_x \sigma_x \label{SO_effe} \quad  
\end{equation}
\\which involves only the $x$-component of spin. In absence of magnetic field all directions are thus equivalent for electron transport, and this explains the lack of dependence on $\theta$ of the I-V curves.

\subsection{Effects of the in-plane magnetic field}
\begin{figure} 
\epsfig{file=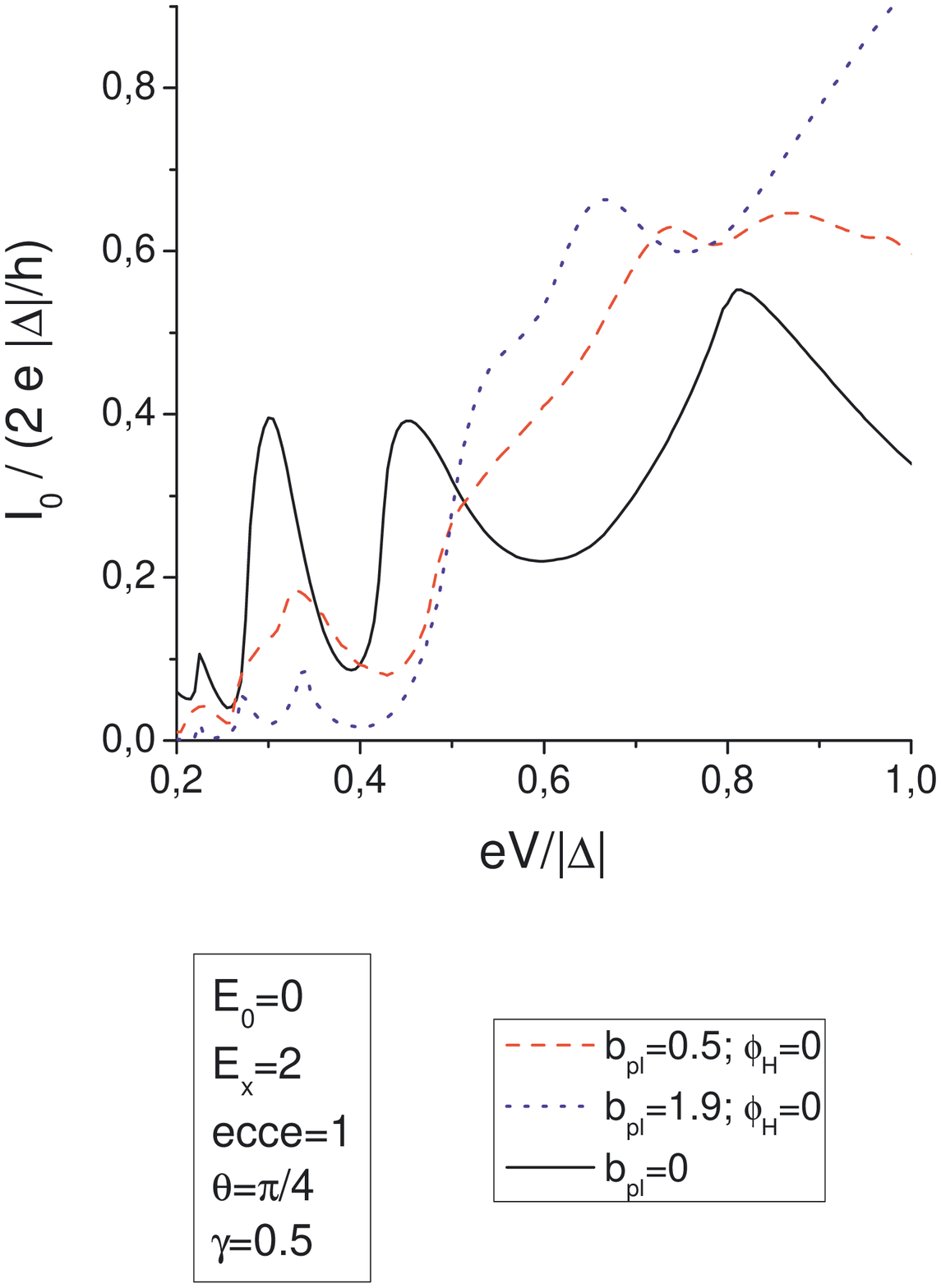,width=8cm,height=8cm,clip=}
\caption{\label{Figura-B-1} (Color online) Effects of the magnetic field on the MAR subgap structure. The case of an isotropic 3-level dot with $E_0=0$, level spacing $\hbar \omega_{x,y}=4 |\Delta|$, spin-orbit parameters $\gamma=0.5$ and $\theta=\pi/4$, and dot-lead linewidths $\Gamma_{\pm,i}=|\Delta|$ ($i=1,2,3$). The solid curve describes the case without magnetic field ($B=0$), whereas the dashed and dotted curves refer to an applied magnetic field along the $x$-axis with intensity $B=0.5$ and $B=1$ respectively. The presence of a magnetic field suppresses the current at low bias and the MAR peaks.}
\end{figure}

In this section we discuss the effect of the in-plane magnetic field, analyzing both the dependence on its intensity~$H$ and on the angle $\phi_H$. Fig.\ref{Figura-B-1} shows the subgap structure for different values of the dimensionless magnetic field
\begin{equation}
B=\frac{g \mu_B H}{|\Delta|} \label{B-def} 
\end{equation}
in an isotropic 3-level quantum dot. As one can see, MAR peaks are suppressed by increasing~$B$, as expected since magnetic field breaks time reversal symmetry and suppresses  the probability that an electron emerging from an Andreev reflection   matches resonance conditions with the appropriate spin direction. \\

\begin{figure} 
\epsfig{file=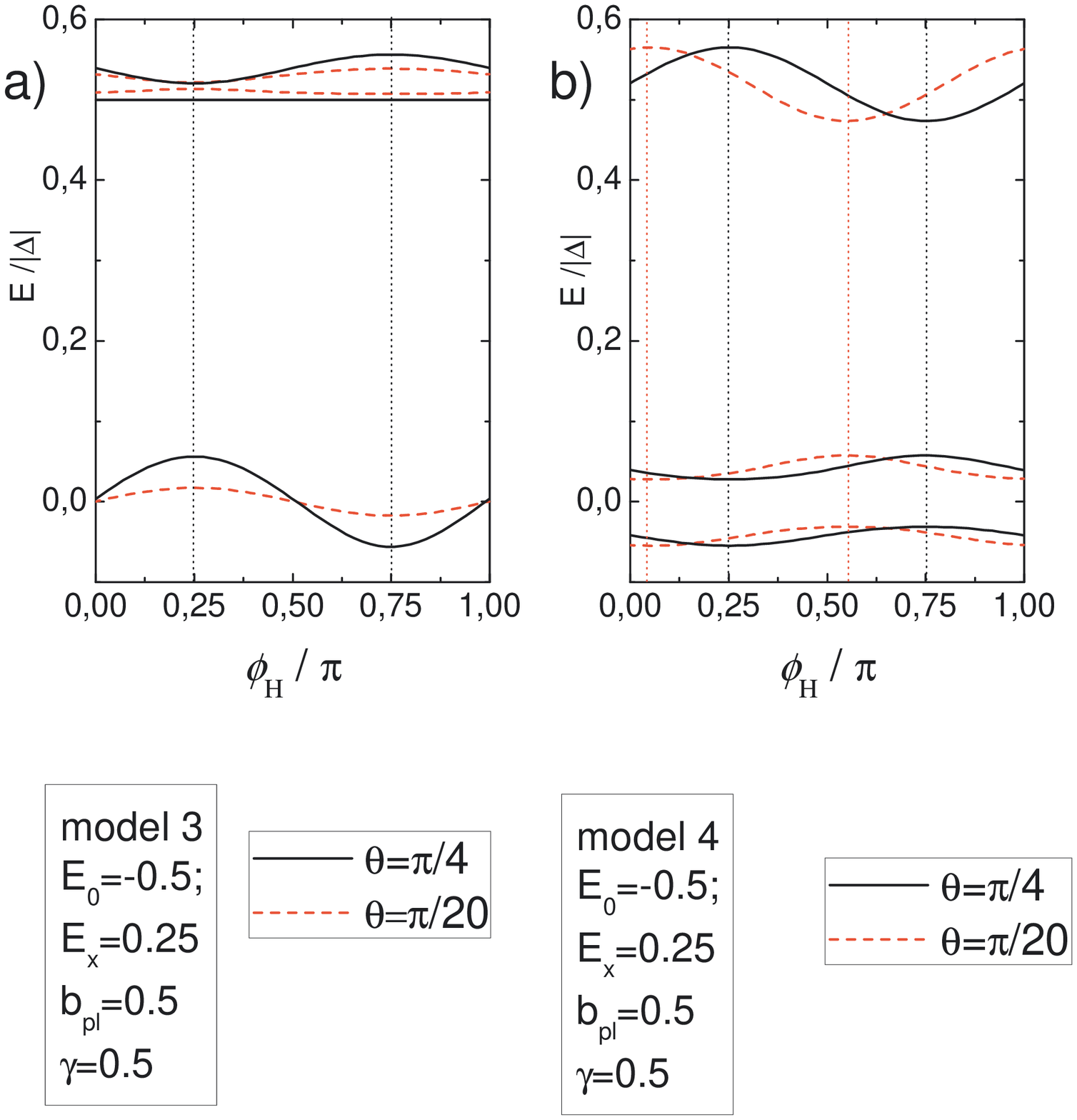,width=8cm,height=6.0cm,clip=}
\caption{\label{Figura-EIGVAL} (Color online) Behavior of (some) dot energy levels as a function of the angle $\phi_H$ of the in-plane magnetic field with dimensionless intensity $B=0.5$ (see Eq.~(\ref{B-def})), and for different values of the spin-orbit Rashba/Dresselhaus angle $\theta$ (solid curves refer to $\theta=\pi/4$ and dashed curves to $\theta=\pi/20$). 
a) case of an isotropic 3-level  dot with $\hbar \omega_x=\hbar \omega_y=|\Delta|/2$; the maxima and minima of   the dot level oscillations are located at $\phi^*_H=\pi (2n+1)/4$ ($n=0,1,2,3$) and are independent of the spin-orbit angle $\theta$, which  determines only the oscillation amplitudes.  b) case of a strongly anisotropic 3-level  dot with $\hbar \omega_x=|\Delta|/2$ and $\hbar \omega_y / |\Delta| \rightarrow \infty$. The maxima and minima of the dot level oscillations are located at the spin-orbit angle $\phi^*_H=\theta \pm n \pi/2$ ($n=0,1,2,3$). Vertical dotted lines are guide for the eye.}
\end{figure}

A more interesting scenario emerges when analyzing the dependence of the subgap structure on the magnetic field angle $\phi_H$. To this purpose, it is worth discussing how the spectrum of the dot varies with $\phi_H$  while keeping the intensity $H$ constant. The shape of the confining potential turns out to play  a crucial role on it. We start with the case of an isotropic quantum dot with three orbital levels (one non-degenerate $s$ level and one doubly degenerate $p$ level). Fig.\ref{Figura-EIGVAL}a shows (some of) the levels of a  dot with  level spacing $\hbar \omega_x=\hbar \omega_y=|\Delta|/2$, in the presence of   a spin-orbit coupling  $\gamma=0.5$ (see Eq.~(\ref{gamma})), and under a magnetic field of intensity $B=0.5$ (see Eq.~(\ref{B-def})).
As one can see, the eigenvalues oscillate as a function of $\phi_H$ and exhibit maxima and minima located at $\phi_H=\pi/4$ and $\phi_H=3\pi/4$. Notice that in this case the spin-orbit angle~$\theta$ only affects the amplitude of the oscillations and not the location of the minima and maxima as a function of~$\phi_H$. 
The origin of this effect boils down to the symmetry of the Hamitonian. Indeed for an isotropic dot the Hamiltonian characterized by a magnetic field with an angle $\phi_H=\pi/4 -\delta\phi_H$ can be mapped into the one with an angle $\phi_H=\pi/4 +\delta\phi_H$ through the unitary transformation
\begin{eqnarray}
\mathsf{V} : \left\{ \begin{array}{lcl} \sigma_x & \leftrightarrow & \sigma_y \\
\sigma_z & \rightarrow & -\sigma_z \\
(x ,p_x) & \leftrightarrow & -(y,p_y)  \end{array}\right.
\end{eqnarray} 
and it therefore exhibits the same spectrum as the latter.
For an isotropic dot, as far as the dependence of I-V curves on the in-plane magnetic field is concerned, the role of $\theta$ is not qualitatively different from the one of spin-orbit strength $\alpha$, and it is thus difficult to extract information about spin-orbit angle by the analysis of the MAR pattern as a function of $\phi_H$. \\
\begin{figure} 
\epsfig{file=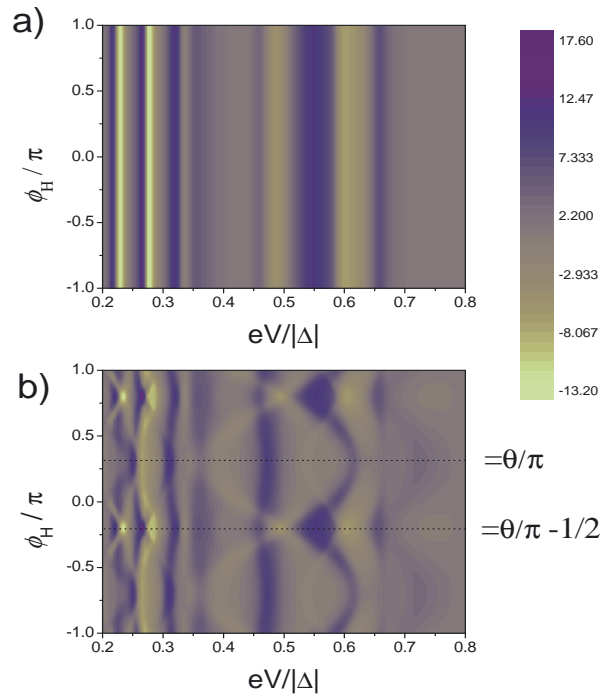,width=8cm,height=9.5cm,clip=}
\caption{\label{Figura-B-3} (Color online) Non-linear conductance $G=dI/dV$ (in units of $2e^2/h$) as a function of the applied bias $V$ (units of $|\Delta|/e$) and of the   angle $\phi_H$ of the in-plane magnetic field with intensity~$B=0.7$ for a strongly anisotropic quantum dot ($\hbar \omega_x=|\Delta|$ and $\hbar \omega_y /|\Delta| \rightarrow \infty$) with two levels, and dot-lead linewidths $\Gamma_{\pm,1,2}=|\Delta|/2$. a) without spin-orbit coupling; b) in the presence of a spin orbit coupling with parameters $\gamma=0.3$ and angle~$\theta=0.3 \pi$. The spin-orbit coupling yields oscillations of the conductance in $\phi_H$ with  minima and maxima given by the spin orbit angle $\theta$ and $\theta-\pi/2$.}
\end{figure}

The situation is quite different for a strongly   anisotropic confining potential $V_\parallel$. The dependence of (some) dot levels is shown in  Fig.\ref{Figura-EIGVAL}b for a 3-level dot with spin orbit coupling $\gamma=0.5$ and magnetic field intensity $B=0.5$. As one can see, the maxima and minima {\it do} depend on the spin-orbit angle, and are precisely located at $\phi^*_H=\theta \pm n \pi/2$ ($n=0,1,2,3$). To understand this effect, we recall that when the anisotropy of the dot confining potential is strong, the electron exhibits a well defined spin orientation which depends on $\theta$; the introduction of a magnetic field simply induces a spin precession around its direction\cite{NOTA-2}. More formally, the  transformation~(\ref{V-theta}) yields an effectively rotated magnetic field
\begin{equation}
\vec{H} \cdot \vec{\sigma} \rightarrow  H   (\cos(\phi_H-\theta+\pi/2) \sigma_x+\sin(\phi_H-\theta+\pi/2) \sigma_y) \label{temp}
\end{equation}
whose direction depends on the spin-orbit angle $\theta$.
The structure of Eq.~(\ref{temp}) shows that, in the presence of an in-plane magnetic field, there are resonances (antiresonances) whenever the angle $\phi_H$ of the physical magnetic field equals $\theta-\pi/2$ ($\theta$), corresponding to an effective magnetic field pointing in the parallel (orthogonal) direction as the spin-orbit term (\ref{SO_effe}). 
A simple example can illustrate this effect: let us consider  the case where only two levels in the $x$ direction are present. In this case the 4 non-degenerate eigenvalues of the dot  can be computed analytically and read  
\begin{eqnarray}
\lefteqn{E=E_0+} & & \label{formula38} \\
& & \frac{\hbar \omega_x}{2} \left(1 \pm \sqrt{1+\gamma^2+b^2 \pm 2 b \sqrt{1+\gamma^2 \sin^2(\theta-\phi_H)}}\right) \nonumber
\end{eqnarray}
where $\gamma$ is the dimensionless spin-orbit coupling (\ref{gamma}), and  $b=g\mu_B H/\hbar \omega$. The resonances between $\theta$ and $\phi_H$ are described by the last oscillatory term in Eq.~(\ref{formula38}). \\

A novel effect thus arises for a strongly anisotropic dot under an in-plane magnetic field: by analyzing  the non-linear conductance $G=dI/dV$ as a function of the bias~$V$ and the magnetic field angle~$\phi_H$, one can see that in the presence of spin-orbit $G$  exhibits maxima and minima located at $\phi_H=\theta$ and $\phi_H=\theta-\pi/2$, i.e. when the magnetic field direction matches the Rashba/Dresselhaus spin-orbit angle. The analysis of the oscillations of the MAR peaks as a function of $\phi_H$ allows to gain the spin-orbit angle $\theta$, as illustrated by the contour plots in Fig.\ref{Figura-B-3} for  quantum dot  with two levels. The upper panel refers to the case without spin-orbit coupling, where no oscillations as a function of~$\phi_H$ are present, since the magnetic field simply determines the preferred spin-direction. The lower panel describes the case of SO coupling $\gamma=0.3$ and $\theta=0.3 \pi$. In this case the magnetic field direction interplays with the SO angle, so that maxima and minima appear at $\phi_H=\theta$ and $\phi_H=\theta-\pi/2$, as indicated on the right of the figure. 

\section{Conclusions}
\label{SecV}
To conclude, we have investigated the effect of Rashba and Dresselhaus spin-orbit couplings on the out-of-equilibrium transport properties of a quantum dot  coupled to two superconductors. We have analyzed how the $I-V$  is affected by the total coupling constant~$\alpha$ (see Eq.~(\ref{alpha})) and by the Rashba/Dresselhaus angle~$\theta$ (see Eq.~(\ref{thetaSO})), discussing also the role of the anisotropy of the confining potential and the in-plane magnetic field for multi-level dots. We have found that, although SO effect prevents electrons tunneling in the dot to have a definite spin-orientation, the MAR subgap structure is not suppressed by SO interaction, due to the conservation of  time-reversal symmetry. The MAR pattern is nevertheless quantitatively modified by $\alpha$, which changes the resonance conditions and the linewidths, affecting the location and the number of the Andreev peaks (see Fig.\ref{Figura-SYM-1}). The role of~$\theta$ is strongly dependent on the shape of the dot confining potential and the magnetic field. In absence of magnetic field, an isotropic dot with equal weights for Rashba and Dresselhaus terms exhibits higher MAR current peaks than the cases with purely Rashba and purely Dresselhaus terms (see Fig.\ref{Figura-SYM-2}), whose I-V curves coincide due to Hamiltonian symmetry. By increasing the dot anisotropy the dependence on $\theta$ vanishes, and the MAR peaks are only affected by the total coupling constant $\alpha$ (see Fig.\ref{Figura-ASYM-1}). When an in-plane magnetic field is applied, the scenario is even richer, especially as a function of the angle $\phi_H$ between the magnetic field direction and the $x$ axis. For an isotropic potential the dot energy levels oscillate as a function of $\phi_H$, with maxima and minima located at  
  $\phi^*_H=\pi (2n+1)/4$ ($n=0,1,2,3$), independent of the spin-orbit angle $\theta$; the latter only determines the oscillation amplitudes and plays a similar role as the coupling $\alpha$. In contrast, in a dot with strongly anisotropic confining potential the maxima and minima are located at  magnetic field angles $\phi^*_H=\theta \pm n \pi/2$ ($n=0,1,2,3$). This enables a direct read-out of the  spin-orbit angle through the inspection of the non-linear conductance as a function of the bias and the angle of the magnetic field $\phi_H$ (see Fig.\ref{Figura-B-3}).
\begin{acknowledgments}
The authors acknowledge  financial support by HYSWITCH and NANOFRIDGE EU Projects, and fruitful discussions with H. Takayanagi, D. Frustaglia, A. Zazunov, V. Pellegrini, F. Giazotto, and R. Egger.
\end{acknowledgments}
\appendix
\section{Evaluation of the Current}
\label{AppA}
In this appendix we show how to obtain Eq.~(\ref{Ip-fin-1}) from Eq.~(\ref{Ip-prel}). As a first step, we observe that, due to the spin-orbit term, states of different orbital levels and opposite spins are coupled, and the Hamiltonian matrix (\ref{HD-mat}) can be diagonalized through a  unitary transformation ${\rm U}$ on the dot level operators
\begin{equation}
d^{}_{n \sigma} =(U_\sigma)_{nj} D^{}_j \hspace{1cm}    j=1,\ldots 2N  \label{U-diag}
\end{equation}
obtaining 
\begin{equation}
\mathcal{H}_{D}=\sum_{i=1}^{2N} \varepsilon_{i} D^\dagger_i D^{}_i
\end{equation}
where $\varepsilon_{i}$ ($i=1,2N$) are the eigenvalues. In Eq.~(\ref{U-diag}) ${\rm U}_\uparrow$ (${\rm U}_\downarrow$) is the $N \times 2N$ upper(lower) submatrix of the unitary transformation ${\rm U}$. Introducing a   block-diagonal $2 \times 4N$ tunneling matrix
\begin{eqnarray}
\hat{T}_{p}   
&= & \! \! \left(
\begin{array}{cc} 
(t^{}_{p,1}   , \ldots ,  t^{}_{p,N} )
\mbox{\rm U}_\uparrow   &   \mbox{\huge{0}}^{}_{}   \\  
& \\
\mbox{\huge{0}}^{}_{}   & -(t^{*}_{p,1}, \ldots  , t^{*}_{p,N}) \mbox{\rm U}^*_\downarrow      \\ 
\end{array} \right) \label{TUN-def}
\end{eqnarray}
with time-dependence given by
\begin{equation}
\hat{T}_{p}(t)= \left( \begin{array}{cc} e^{i p \omega_V t/2} & 0 \\ 0 & e^{-i p \omega_V t/2} \end{array} \right) \cdot \hat{T}_{p}(0) \label{T-time}
\end{equation}
one can rewrite Eq.~(\ref{Htun}) in a Nambu notation as follows
\begin{eqnarray}
\mathcal{H}_{tun,p} &=&        \sum_{\mathbf{k}}     
\left( 
c^{\dagger}_{\mathbf{k},p, \uparrow}  ,
c^{}_{-\mathbf{k},p, \downarrow} 
 \right) 
 \cdot  \hat{T}_{p}(t) \cdot
\left( 
\begin{array}{c}
D_{1} \\ 
\vdots \\
D_{2N} \\
D^\dagger_{1} \\ 
\vdots \\
D^\dagger_{2N}
\end{array}
\right)  +    \mbox{h.c.}
 \nonumber  
\end{eqnarray}
Similarly, the current (\ref{Ip-prel}) can be rewritten as
\begin{eqnarray}
\lefteqn{I_p(t) =} & & \label{Ip-pre}\\
&=&-\frac{2e}{\hbar} \Re \sum_{\mathbf{k}} {\rm Tr}_{4N} \left\{   (\sigma_z)_{4N} \left[ \mathsf{T}^\dagger_{p}(t) \mathsf{G}_{\mathbf{k}p,0}(t,t)  \right]^{+-} \right\}  \nonumber
\end{eqnarray}
where
\begin{equation}
\mathsf{T}^\dagger_{p} = \left( \begin{array}{cc} \hat{T}^\dagger_{p} & 0 \\ 0 & -\hat{T}^\dagger_{p} \end{array} \right) \quad,    \label{Textdag-def}
\end{equation}
and 
\begin{equation}
\mathsf{G}_{\mathbf{k}p,0}=  \left( \begin{array}{cc} G_{\mathbf{k}p,0}^{++} & G_{\mathbf{k}p,0}^{+-} \\ G_{\mathbf{k}p,0}^{-+} & G_{\mathbf{k}p,0}^{--} \end{array} \right)  \label{Gkp0-kel}
\end{equation}
denotes the lead-dot Green functions with entries
\begin{eqnarray}
\lefteqn{iG^{\eta_1 \eta_2}_{\mathbf{k}p,0}(t_1,t_2)= } & &   \label{Gkp0-def}\\
& & \! \! \! \! \Bigg\langle \!
\left( 
\begin{array}{c}
{c^{}}^{(\eta_1)}_{\mathbf{k} p  \uparrow} \! \\ \\
{c^{\dagger}}^{(\eta_1)}_{-\mathbf{k} p  \downarrow} \!
\end{array} \right) \! (t_1) \! \cdot  \!
\left( 
{D^\dagger}^{(\eta_2)}_{1} , 
\cdot \cdot,
{D^\dagger}^{(\eta_2)}_{2N} ,
D^{(\eta_2)}_{1} ,
\cdot \cdot ,
D^{(\eta_2)}_{2N}
\right)(t_2)  \Bigg\rangle   \nonumber 
\end{eqnarray}
with $\eta_{1,2}=\pm$  Keldysh labels. Exploiting Keldysh-Dyson equation
\begin{equation}
\mathsf{G}_{\mathbf{k}p,0}(t,t)=\frac{1}{\hbar}  \int_{-\infty}^{\infty} \, \mathsf{g}_{\mathbf{k}p,\mathbf{k}p} (t,t^\prime) \mathsf{T}_{\mathbf{k}p}(t^\prime) \, \mathsf{G}(t^\prime,t) \, \, dt^\prime \label{Dyson}
\end{equation} 
one can express $\mathsf{G}_{\mathbf{k}p,0}$ in terms of the  Green function $\mathsf{g}_{\mathbf{k}p,\mathbf{k}p}$ of the $p$-lead and the Green function $\mathsf{G}$ of the dot. The latter are defined similarly to Eq.~(\ref{Gkp0-kel}) with entries
\begin{equation}
ig_{\mathbf{k}p,\mathbf{k}p}^{\eta_1 \eta_2}(t_1,t_2) =   \Bigg\langle \left( 
\begin{array}{c}
c^{(\eta_1)}_{\mathbf{k} p  \uparrow} \\ \\
{c^{\dagger}}^{(\eta_1)}_{-\mathbf{k} p  \downarrow}  
\end{array} \right)(t_1) \left( 
\begin{array}{cc}
{c^{\dagger}}^{(\eta_2)}_{\mathbf{k} p \uparrow}  &
{c^{}}^{(\eta_2)}_{-\mathbf{k} p  \downarrow}
\end{array} \right)(t_2) \Bigg\rangle,
\end{equation}
and
\begin{eqnarray}
\lefteqn{ iG^{\eta_1 \eta_2}(t_1,t_2)= } & & \\ 
& & \Bigg\langle 
\left( 
\begin{array}{c}
D_{1}  \\ 
\vdots \\
D_{2N}  \\
D^\dagger_{1}  \\ 
\vdots \\
D^\dagger_{2N}
\end{array}
\right)(t_1) \left( 
\begin{array}{cccccc}
D^\dagger_{1}  & 
\ldots &
D^\dagger_{2N}  &
D^{}_{1}  &
\ldots &
D^{}_{2N}
\end{array}
\right)(t_2)  \Bigg\rangle \nonumber
\end{eqnarray}
respectively. Inserting Eq.~(\ref{Dyson}) into Eq.~(\ref{Ip-pre}) and recalling that  the self-energy due to the $p$-th lead reads
\begin{equation}
\mathsf{\Sigma}_{p}(t_1,t_2)=\frac{1}{\hbar^2} \sum_{\mathbf{k}}  \mathsf{T}^\dagger_{\mathbf{k}p}(t_1)   \, \mathsf{g}_{\mathbf{k}p,\mathbf{k}p} (t_1,t_2) \mathsf{T}_{\mathbf{k}p}(t_2) \,   \, \, \label{SELF}
\end{equation} 
one obtains Eq.~(\ref{Ip-fin-1}).
\section{Discrete Fourier Transform}
\label{AppB}
In this appendix we sketch the calculation that leads to the evaluation of $\mathsf{\Sigma}_{p}(n_1,n_2;\omega)$ and $\mathsf{G}(n_1,n_2;\omega)$ appearing in the current formula (\ref{Ip-fin}). One first introduces\cite{Alfredo} a discrete Fourier transform $f(n,m;\omega)$ of an arbitrary two time arguments function $f(t_1,t_2)$, defined through the relation
\begin{eqnarray}
\lefteqn{f(t_1,t_2)=} \label{AFT} & & \\
& & \sum_{n,m=-\infty}^{\infty} \int_F \frac{d\omega}{2\pi} \, e^{-i (\omega+n \omega_V)t_1} \, e^{i (\omega+m \omega_V)t_2} \, f(n,m;\omega) \nonumber 
\end{eqnarray} 
where $F$ is the fundamental domain introduced in Eq.~(\ref{FUND-DOM}).
The inversion formula reads
\begin{eqnarray}
\lefteqn{2 \pi \delta(\omega_1-\omega_2) \, f(n_1,n_2;\omega_1)=}  \label{AFT-inv}& & \\
& &=\int_{-\infty}^{\infty} \!\! dt \int_{-\infty}^{\infty}\!\! dt^\prime  \, f(t,t^\prime) \,  e^{i (\omega_1+n_1 \omega_V)t} \, e^{-i (\omega_2+n_2 \omega_V)t^\prime} \, \, \nonumber
\end{eqnarray}
\\where $\omega_1, \omega_2 \in F$. For a function that depends only on time differences $f(t_1,t_2)=f(t_1-t_2)$, the discrete Fourier transform reads
\begin{equation}
f(n_1,n_2;\omega)=\delta_{n_1,n_2} \, \tilde{f}(\omega+n_1 \omega_V)
\end{equation}
where $\tilde{f}(\omega)$ is the usual Fourier Transform.\\

The function $\mathsf{\Sigma}_{p}(n_1,n_2;\omega)$ is thus  obtained by inserting Eq.~(\ref{SELF})  as $f$ into the right-hand side of Eq.~(\ref{AFT-inv}), and making use of Eq.~(\ref{T-time}) and of the definition
\begin{equation}
\mathsf{g}_{p}(\omega) = \int_{-\infty}^{+\infty} dt \, e^{i \omega t} \sum_{\mathbf{k}}  \, \mathsf{g}_{\mathbf{k}p,\mathbf{k}p} (t)  
\end{equation}
obtaining
\begin{widetext}
\begin{eqnarray}
\mathsf{\Sigma}_{p}(n_1,n_2;\omega)   
&=& \frac{1}{\hbar^2} \, \mathsf{T}^\dagger_{p}(0)   
\left( 
\begin{array}{ccc}   
 \delta_{n_2,n_1} [\mathsf{g}_{p}(\omega+n_1 \omega_V -\frac{p \omega_V}{2})]_{11}   
&  & 
\delta_{n_2,n_1-p}  [\mathsf{g}_{p}(\omega+n_1 \omega_V -\frac{p \omega_V}{2})]_{12} 
\\ & & \\ & & \\ 
\delta_{n_2,n_1+p} [\mathsf{g}_{p}(\omega+n_1 \omega_V +\frac{p \omega_V}{2})]_{21}    
&  & 
 \delta_{n_2,n_1}  [\mathsf{g}_{p}(\omega+n_1 \omega_V +\frac{p \omega_V}{2})]_{22}    
\end{array} \right)   \mathsf{T}_{p}(0)    \label{SELF-omega}
\end{eqnarray}
where $[ \ldots ]_{ij}$ ($i,j=1,2$) denote  the entries in the $2$-dimensional Nambu space of the leads.
A standard calculation yields\cite{Kopnin}
\begin{eqnarray}
\mathsf{g}_{p}(\omega)= \pi \rho(\varepsilon_F) 
\left[  h_1(\omega)  \left( \begin{array}{cc} 1 & 0 \\ 0 & -1 \end{array} \right)_{\rm K} \, + i h_2(\omega) \left( \begin{array}{cc} 2 f_p(\omega)-1 & 2 f_p(\omega) \\ 2f_p(\omega)-2 & 2f_p(\omega)-1 \end{array} \right)_{\rm K} \right] \otimes \left( \begin{array}{cc} 1 & -\frac{\Delta_p}{\hbar \omega} \\ -\frac{\Delta_p}{\hbar \omega} & 1 \end{array} \right)_{\rm N} \label{g-lead-fin}
\end{eqnarray} 
\end{widetext}
where $f_p$ denotes the Fermi function of the $p$-the lead, and
\begin{equation}
h_1(\omega)=\frac{-\hbar \omega \Theta(|\Delta_p|-|\omega|)}{\sqrt{|\Delta_p|^2-\omega^2}} \quad,
\end{equation}
\begin{equation}
h_2(\omega)=  \frac{|\hbar \omega| \Theta(|\omega|-|\Delta_p|)}{\sqrt{\omega^2-|\Delta_p|^2} } \quad,
\end{equation}
with $\Theta$ being the Heaviside function. In Eq.~(\ref{g-lead-fin}) the symbols $( \ldots )_{\rm N}$ and ,   $( \ldots )_{\rm K}$ denote matrices acting on the lead Nambu and Keldysh space respectively.\\

Finally the dot Green function $\mathsf{G}$ can be evaluated by means of the second Dyson equation (\ref{Dyson-2}), where $\mathsf{G}_0$ describes the  green function of the isolated dot. Explicitly $\mathsf{G}^{-1}_0(n_1,n_2;\omega)=\delta_{n_1,n_2} \mathsf{G}^{-1}_0(\omega+n_1 \omega_V)$, where
\begin{equation}
\mathsf{G}^{-1}_0(\omega)= \omega \, {\mathbb{I}}_{4N} - \sigma_z \otimes \mbox{diag}( \varepsilon_1 , \ldots,  \varepsilon_{2N} )
\end{equation}
with ${\mathbb{I}}_{4N}$ the $4N \times 4N$ identity matrix.


\begin{thebibliography}{99}
\bibitem{NOTA} If the spin-orbit coupling is much bigger that the dot level spacing this arguments breaks obviously down. Indeed in Ref.\onlinecite{Luca} it has been predicted that SO effects can be observed also for a single-level quantum dot, provided that a magnetic field is present and that the SO coupling   is sufficiently strong. Our results are instead limited to SO coupling not higher than the dot level spacing and, in this regime, our approach agrees with the one of Ref.\onlinecite{Luca}.
\bibitem{NOTA-2} Notice that in the case of a perfectly isotropic dot (Fig.\ref{Figura-EIGVAL}a) the location of maxima and minima at $\pi/4$ and $3 \pi/4$ precisely corresponds to those particular spin-orbit angles~$\theta$ for which the electron spin has a well defined direction. \\
\end{thebibliography}
\end{document}